\documentclass[12pt]{article}

\usepackage{graphicx}
\usepackage{amsmath,amssymb,url}

\allowdisplaybreaks[1]

\begin{document}


\def\TODO#1{ {\bf ($\clubsuit$ #1 $\clubsuit$)} }


\baselineskip 0.7cm

\begin{titlepage}

\hfill UT--12--12

\vskip 1.35cm
\begin{center}
{\large \bf
Mass Insertion Formula for Chargino Contribution \\
to $\Delta B = 1$ Wilson Coefficients \\
and its Application to CP Asymmetries of $B \to K\pi$
}
\vskip 1.2cm
Motoi Endo and Takahiro Yoshinaga
\vskip 0.4cm

Department of Physics, University of Tokyo, Tokyo 113-0033, Japan

\vskip 1.5cm

\abstract{
The comprehensive mass insertion formula of the $b \to s \bar q q$ transitions was provided for the chargino contributions. The method clarifies physics behind the calculations of the supersymmetric contributions to FCNCs. The result was applied to the analysis of the difference of the CP asymmetries of the $B \to K\pi$ decays. It is shown that the supersymmetry is not responsible for the discrepancy. 
}
\end{center}
\end{titlepage}

\setcounter{page}{2}


\section{Introduction}

The CKM picture of the standard model (SM) has been confirmed by experimental and theoretical developments \cite{Asner:2010qj,Laiho:2012ss,Lenz:2012az,Bona:2009cj}. Although a lot of the experimental results of flavor-changing neutral currents (FCNCs) are consistent with the SM, there are processes whose measurements deviate from the SM prediction \cite{Buras:2012ts}. Some of them have been reported in the $b-s$ transition processes. Definitely, it is too early to conclude the anomalies due to physics beyond the SM unless uncertainties especially from hadronic contributions are reduced. Nonetheless, it is important to explore the new physics (NP) in the flavor sector, because the FCNCs are sensitive to flavor violations in TeV scale. 

Supersymmetry (SUSY) is one of the most attractive extensions of the SM. It involves new flavor structures due to scalar partners of the SM matters. They contribute to the $b-s$ transitions by gluino and chargino diagrams \cite{Gabbiani:1996hi}. There are two methods to describe them: i) the loop diagrams are evaluated without approximations \cite{Inami:1980fz}, and ii) they are expanded by small parameters (see, e.g., \cite{Gabbiani:1996hi}). The former provides full evaluations, while the latter has virtues that the contributions can be understood clearly. In fact, physics behind the former calculations becomes clearer, and cancellations which appear in the former method such as the GIM mechanism are performed automatically. In the literature, the latter description, called the mass insertion (MI) approximation, has been discussed for the gluino contributions. The MI formula for the chargino contributions to $b \to s \ell\ell$ is also given in Ref.~\cite{Lunghi:1999uk}, though some typos were recently reported \cite{Behring:2012mv}. That for $b \to s \bar q q$ are found in Ref.~\cite{Chakraverty:2003uv,Gabrielli:2004yi}. However, some contributions are not included, as will be shown later. In this paper, the comprehensive chargino contributions to the Wilson coefficients of $b \to s \bar q q$ will be provided in the MI approximation. 

The analysis will be applied to the $K\pi$ puzzle. The $B \to K\pi$ decay is dominated by the penguin contribution and sensitive to the NP (see, e.g., \cite{Mishima:2011qy} for a review). Currently, the measurement of the CP asymmetry of the $B^0 \to K^+ \pi^-$ decay is different from that of $B^\pm \to K^\pm \pi^0$ at more than the 5$\sigma$ level \cite{Asner:2010qj}. In the SM, such a large deviation is unexpected (see \cite{Mishima:2011qy} for a summary of the theoretical predictions), which is called the $K\pi$ puzzle. In the topological decomposition of the amplitudes \cite{Gronau:1994rj}, the difference originates in the color-suppressed tree ($C$) and electroweak penguin ($P_{\rm ew}$) contributions. The experimental results may indicate enhancement of $C$ and/or $P_{\rm ew}$ by NP contributions. 

The SUSY solution to the puzzle by means of the chargino contributions has been discussed in \cite{Khalil:2005qg,Khalil:2006hj,Khalil:2009zf}. In contrast to the gluino contributions which have already been constrained \cite{Imbeault:2008ge}, the authors insisted that the chargino can induce a large CP asymmetry of $B \to K\pi$. However, it was reanalyzed in \cite{Hofer:2010ee} with the full loop calculations, i.e., i) in the above category, and found to be less significant. This conclusion will be confirmed by the MI approach. Also, the contributions with the squark mixing of $(\delta^u_{LR})_{32}$ are analyzed, which were discussed in \cite{Khalil:2005qg,Khalil:2006hj,Khalil:2009zf}. They can be included by turning on the charm Yukawa coupling and will be shown to be too small to solve the $K\pi$ puzzle explicitly by using the MI formula.

This paper is organized as follows. The Wilson coefficients of the chargino contributions as well as those of the gluino are provided in Sec.~\ref{sec:wilson}. They are applied to the CP asymmetries of $B \to K\pi$ in Sec.~\ref{sec:btokpi}. The results are summarized in Sec.~\ref{sec:conclusion}.


\section{Wilson Coeffients}
\label{sec:wilson}

The $b \to s \bar q q$ transitions are represented by the following effective Hamiltonian \cite{Buchalla:1995vs},
\begin{align}
 H_{\mathrm{eff}} &= \frac{4G_F}{\sqrt{2}}\sum_{p=u,c} \lambda_p
 \left[ \sum_{i=1,2} C_i \mathcal{O}_i^p 
 + \sum^{10}_{i=3}C_i \mathcal{O}_i + C_{7\gamma } \mathcal{O}_{7\gamma }
 + C_{8G} \mathcal{O}_{8G} + (L \leftrightarrow R) \right] + {\rm{h.c.}},
 \label{eq:Heff}
\end{align}
by integrating out heavy degrees of freedom, where $q$ is the $u$ or $d$ quark. Here, $\lambda_p$ is defined by the CKM matrix as $\lambda_p = V_{pb}V_{ps}^*$. The operators are defined as \footnote{
There are additional operators in the effective Hamiltonian up to the dimension six such as those with the scalar or tensor Lorentz structure. In this paper, we are interested in the processes that are not suppressed in the chiral limit. Since the extra  operators involve the chirality flip of the quarks, they are irrelevant, and the operators in (\ref{eq:operators}) are enough. 
}
\begin{align}
 &\mathcal{O}_1^p =(\bar{p}^{\alpha }_L \gamma^{\mu } b^{\beta }_L) (\bar{s}^{\beta }_L \gamma_{\mu } p^{\alpha }_L), &
 &\mathcal{O}_2^p =(\bar{p}^{\alpha }_L \gamma^{\mu } b^{\alpha }_L) (\bar{s}^{\beta }_L \gamma_{\mu } p^{\beta }_L), \notag \\
 &\mathcal{O}_3 = (\bar{s}^{\alpha }_L \gamma^{\mu } b^{\alpha }_L) 
 \sum_{q=u,d} (\bar{q}^{\beta }_L \gamma_{\mu } q^{\beta }_L), &
 &\mathcal{O}_4 = (\bar{s}^{\alpha }_L \gamma^{\mu } b^{\beta }_L) 
 \sum_{q=u,d}  (\bar{q}^{\beta }_L \gamma_{\mu } q^{\alpha }_L), \notag \\
 &\mathcal{O}_5 = (\bar{s}^{\alpha }_L \gamma^{\mu } b^{\alpha }_L) 
 \sum_{q=u,d}  (\bar{q}^{\beta }_R \gamma_{\mu } q^{\beta }_R), &
 &\mathcal{O}_6 = (\bar{s}^{\alpha }_L \gamma^{\mu } b^{\beta }_L) 
 \sum_{q=u,d}  (\bar{q}^{\beta }_R \gamma_{\mu } q^{\alpha }_R), \notag \\
 &\mathcal{O}_7 = (\bar{s}^{\alpha }_L \gamma^{\mu } b^{\alpha }_L) 
 \sum_{q=u,d}  \frac{3}{2}e_q (\bar{q}^{\beta }_R \gamma_{\mu } q^{\beta }_R), &
 &\mathcal{O}_8 = (\bar{s}^{\alpha }_L \gamma^{\mu } b^{\beta }_L) 
 \sum_{q=u,d}  \frac{3}{2}e_q (\bar{q}^{\beta }_R \gamma_{\mu } q^{\alpha }_R), \notag \\
 &\mathcal{O}_9 = (\bar{s}^{\alpha }_L \gamma^{\mu } b^{\alpha }_L) 
 \sum_{q=u,d}  \frac{3}{2}e_q (\bar{q}^{\beta }_L \gamma_{\mu } q^{\beta }_L), &
 &\mathcal{O}_{10} = (\bar{s}^{\alpha }_L \gamma^{\mu } b^{\beta }_L) 
 \sum_{q=u,d}  \frac{3}{2}e_q (\bar{q}^{\beta }_L \gamma_{\mu } q^{\alpha }_L), \notag \\
 &\mathcal{O}_{7\gamma } = \frac{e}{16\pi^2}(\bar{s}_L \sigma_{\mu \nu }b_R) F^{\mu \nu } , &
 &\mathcal{O}_{8G} = \frac{g_s}{16\pi^2} (\bar{s}_L \sigma_{\mu \nu } b_R) G^{\mu \nu },
 \label{eq:operators}
\end{align}
where the color indices are denoted by $\alpha$ and $\beta$, and the electromagnetic and strong coupling constants are $e$ and $g_s$, respectively. In the electroweak (EW) penguins, $C_{i = 7 - 10}$, $e_q$ is the electromagnetic charge of the quark, $q$. The above operators are often called the left-handed currents, while the chirality-flipped operators and their Wilson coefficients, $\mathcal{\tilde O}_i$ and $\tilde C_i$, are called the right-handed currents. 

Focusing on the transitions between the second and third generations, there are eight types of the squark mixings, which are classified by the chirality index of the squarks and whether the mixing belongs to the up- or down-type squarks. Let us consider the superCKM basis, where the quark supermultiplets are rotated such that the quark mass matrices are diagonalized. Off-diagonal components of the squark mass matrices, $-\mathcal{L} = \tilde{b}_i^* (\Delta m_{\tilde{d}}^2 )_{ij}\tilde{s}_j + \tilde{t}_i^* (\Delta m_{\tilde{u}}^2 )_{ij}\tilde{c}_j$, are parametrized as $(\delta_{ij}^q)_{32} = (\Delta m_{\tilde{q}}^2 )_{ij}/\bar m_{\tilde q}^2$, where $i, j$ denote the chirality of the squark, $L$ and $R$, and $\bar m_{\tilde q}^2$ is a diagonal component of the squark mass matrix. Thus, the SUSY contributions are represented by the eight MI parameters,
\begin{align}
 (\delta_{LL}^d)_{32},~~(\delta_{RL}^d)_{32},~~(\delta_{RR}^d)_{32},~~(\delta_{LR}^d)_{32},\notag \\
 (\delta_{LL}^u)_{32},~~(\delta_{RL}^u)_{32},~~(\delta_{RR}^u)_{32},~~(\delta_{LR}^u)_{32}.
\end{align}
In this paper, the soft squark masses are simply set to be same except for the right-handed stop, and $\bar m_{\tilde q}^2$ is chosen to be this common mass. For the down-type squark mixings, it is convenient to use their Hermitian conjugation, $(\delta_{ij}^d)_{23} = (\delta_{ji}^d)_{32}^*$.

When the down-type squark mixings are present, the gluino--squark diagrams dominate the SUSY contributions to $b \to s \bar q q$. The Wilson coefficients are obtained as \cite{Gabbiani:1996hi,Endo:2004xt}:
\begin{align}
 C_3 &= \frac{\sqrt{2}\alpha_s^2}{4G_F\lambda_tm_{\tilde{q}}^2}(\delta^d_{LL})_{23}
 \left[-\frac{1}{9}B_1(x)-\frac{5}{9}B_2(x)-\frac{1}{18}P_1(x)-\frac{1}{2}P_2(x) \right], \notag \\
 C_4 &= \frac{\sqrt{2}\alpha_s^2}{4G_F\lambda_tm_{\tilde{q}}^2}(\delta^d_{LL})_{23}
 \left[-\frac{7}{3}B_1(x)+\frac{1}{3}B_2(x)+\frac{1}{6}P_1(x)+\frac{3}{2}P_2(x) \right], \notag \\
 C_5 &= \frac{\sqrt{2}\alpha_s^2}{4G_F\lambda_tm_{\tilde{q}}^2}(\delta^d_{LL})_{23}
 \left[\frac{10}{9}B_1(x)+\frac{1}{18}B_2(x)-\frac{1}{18}P_1(x)-\frac{1}{2}P_2(x) \right], \notag \\
 C_6 &= \frac{\sqrt{2}\alpha_s^2}{4G_F\lambda_tm_{\tilde{q}}^2}(\delta^d_{LL})_{23}
 \left[-\frac{2}{3}B_1(x)+\frac{7}{6}B_2(x)+\frac{1}{6}P_1(x)+\frac{3}{2}P_2(x) \right], \notag \\
 C_{7\gamma } &= \frac{\sqrt{2}\alpha_s\pi }{4G_F\lambda_tm_{\tilde{q}}^2}\bigg[(\delta^d_{LL})_{23}
 \bigg( -\frac{16}{9}M_3(x) + \mu_H \tan \beta \frac{m_{\tilde{g}}}{m_{\tilde{q}}^2}\frac{16}{9}M_a(x)\bigg)
 +(\delta^d_{LR})_{23}\frac{m_{\tilde{g}}}{m_b}\frac{8}{3}M_1(x)\bigg], \notag \\
 C_{8G} &= \frac{\sqrt{2} \alpha_s\pi }{4 G_F\lambda_tm_{\tilde{q}}^2}\bigg[(\delta^d_{LL})_{23}
 \bigg\{ \bigg( -\frac{2}{3}M_3(x) -6M_4(x)\bigg) 
 + \mu_H \tan \beta \frac{m_{\tilde{g}}}{m_{\tilde{q}}^2}\bigg( \frac{2}{3}M_a(x)
 +6M_b(x) \bigg) \bigg\} \notag \\
 &~~~~~~~~~~~~~~~~~~~
 +(\delta^d_{LR})_{23}\frac{m_{\tilde{g}}}{m_b}\bigg( \frac{1}{3}M_1(x) +3M_2(x) \bigg) \bigg]. 
 \label{eq:gluino}
\end{align}
where $B_{1,2} P_{1,2}, M_{1-4}$ and $M_{a,b}$ are the loop functions (see App.~\ref{app:functions}) with $x = m_{\tilde{g}}^2/m_{\tilde{q}}^2$. Here and hereafter, the Yukawa coupling of the strange quark is neglected. The contributions to the right-handed currents are obtained by flipping the chirality such as $(\delta^d_{LL})_{23} \to (\delta^d_{RR})_{23}$ and $(\delta^d_{LR})_{23} \to (\delta^d_{RL})_{23}$. It  is noticed that the magnetic dipole terms, $C_{7\gamma}$ and $C_{8G}$, have two types of the contributions for $(\delta^d_{LL})_{23}$ are $(\delta^d_{RR})_{23}$. It is easy to see that the $\tan\beta$-enhanced one is usually dominant, which is often called the double MI contribution or the induced $LR$ mixing. 

For the four-quark operators, the gluon box and the gluon penguin contributions are included in (\ref{eq:gluino}). There are additionally the EW penguins by mediating the photon \cite{Khalil:2005qg},
\begin{align}
 C_7 = C_9 = \frac{\sqrt{2}\alpha_s \alpha_e}{4G_F\lambda_tm_{\tilde{q}}^2}(\delta^d_{LL})_{23}
 \left[ -\frac{16}{27}P_1(x) \right],
 \label{eq:gluino-photon}
\end{align}
while those to $C_8$ and $C_{10}$ are small. The contributions to $\tilde C_i$ are obtained by flipping the chirality. The following Z-penguin contributions are also added to the coefficients,
\begin{align}
 C_3 &= -\frac{\alpha _s}{18\pi \lambda _t m_{\tilde{q}}^2}m_b\mu _H \tan \beta \bigg[ (\delta ^d_{LL})_{23} \frac{\mu _H \tan \beta }{m_{\tilde{q}}^2}Z_b(x)+ (\delta ^d_{LR})_{23}Z_a(x) \bigg],\notag \\
 C_7 &= -\frac{2\alpha _s}{9\pi  \lambda _t m_{\tilde{q}}^2}\sin ^2\theta _Wm_b\mu _H \tan \beta  \bigg[ (\delta ^d_{LL})_{23} \frac{\mu _H \tan \beta }{m_{\tilde{q}}^2}Z_b(x)+ (\delta ^d_{LR})_{23}Z_a(x) \bigg],\notag \\
 C_9 &= \frac{2\alpha _s}{9\pi  \lambda _t m_{\tilde{q}}^2}\cos ^2\theta _Wm_b\mu _H \tan \beta  \bigg[ (\delta ^d_{LL})_{23} \frac{\mu _H \tan \beta }{m_{\tilde{q}}^2}Z_b(x)+ (\delta ^d_{LR})_{23}Z_a(x) \bigg], \notag \\
 \tilde{C}_5 &= \frac{\alpha _s}{9\pi  \lambda _t m_{\tilde{q}}^2}m_b\mu _H \tan \beta \bigg[ (\delta ^d_{RR})_{23} \frac{\mu _H \tan \beta }{m_{\tilde{q}}^2}Z_b(x)+ (\delta ^d_{RL})_{23}Z_a(x) \bigg], \notag \\
 \tilde{C}_7 &= -\frac{4\alpha _s}{9\pi  \lambda _t m_{\tilde{q}}^2}\cos ^2\theta _Wm_b\mu _H \tan \beta \bigg[ (\delta ^d_{RR})_{23} \frac{\mu _H \tan \beta }{m_{\tilde{q}}^2}Z_b(x)+ (\delta ^d_{RL})_{23}Z_a(x) \bigg], \notag \\
 \tilde{C}_9 &= \frac{4\alpha _s}{9\pi  \lambda _t m_{\tilde{q}}^2}\sin ^2\theta _Wm_b\mu _H \tan \beta \bigg[ (\delta ^d_{RR})_{23} \frac{\mu _H \tan \beta }{m_{\tilde{q}}^2}Z_b(x)+ (\delta ^d_{RL})_{23}Z_a(x) \bigg].
\label{eq:gluino-Z}
\end{align}
The loop functions, $Z_a$ and $Z_b$, are found in App.~\ref{app:functions}. The Z penguins with $(\delta^d_{LL})_{23}$ and $(\delta^d_{RR})_{23}$ can be comparable to (\ref{eq:gluino-photon}) when $\tan\beta$ is large, though they are proportional to $m_b$. Lastly, the box diagram contributions to $C_7$ and $C_8$ can be finite for $(\delta^d_{LL})_{23}$ and $(\delta^d_{RR})_{23}$, when there is a large difference between the sup and sdown masses. Since they are smaller than (\ref{eq:gluino-photon}) by $\sim (m_{\tilde d_R}^2 - m_{\tilde u_R}^2)/m_{\tilde q}^2$, they are discarded in this paper. 

The up-type squark mixings appear in the chargino contributions to $b \to s \bar q q$. The charginos consist of the Wino and Higgsino, which mix to each other by the electroweak symmetry breaking. Since the weak boson masses are much smaller than the SUSY scale, the chargino amplitudes can be expanded with respect to $m_{W,Z}/m_{\rm soft}$ as well as $(\delta_{ij}^q)_{32}$. The Wilson coefficients of the four-quark operators are classified as
\begin{align}
 C_3 &= \frac{1}{6}C_Z -\frac{1}{3}C_g + \frac{1}{2}C_{B_u} + C_{B_d},~~~
 C_4 = C_g,~~~
 C_5 = -\frac{1}{3}C_g,~~~
 C_6 = C_g ,\notag \\
 C_7 &= C_{\gamma } + \frac{2}{3}\sin^2 \theta_W C_Z,~~~
 C_9 = C_{\gamma } - \frac{2}{3}\cos^2 \theta_W C_Z + C_{B_u} - C_{B_d},\notag 
\end{align}
where $C_g$ represents the gluon penguins, $C_{\gamma }$ the photon penguins, $C_Z$ the Z penguins, and $C_{B_u}$ and $C_{B_d}$ the box diagrams. In the MI approximation, they become
\begin{align}
 C_g &= \frac{\alpha_s}{27\pi \lambda_t}\frac{M_W^2}{m^2_{\tilde{q}}}\bigg[ (\delta^u_{LL})_{32}P^{LL}_g(x_{\tilde{W}}) 
 \notag\\&\quad
 + (\delta^u_{RL})_{32} \bigg\{ \frac{m_t M_2}{m^2_{\tilde{q}}}P^{RL1}_g(x_{\tilde{W}}, x_{\mu }, x_{\tilde{t}_R}) 
 + \frac{m_t A_t}{m^2_{\tilde{q}}}P^{RL2}_g(x_{\tilde{W}}, x_{\tilde{t}_R}) \bigg\} \bigg],
 \notag\\ 
 C_{\gamma} &= 
 \frac{\alpha_e}{27\pi \lambda_t}\frac{M_W^2}{m^2_{\tilde{q}}}\bigg[ (\delta^u_{LL})_{32}P^{LL}_{\gamma }(x_{\tilde{W}}) 
 \notag\\&\quad
 + (\delta^u_{RL})_{32} \bigg\{ \frac{m_t M_2}{m^2_{\tilde{q}}}P^{RL1}_{\gamma }(x_{\tilde{W}}, x_{\mu }, x_{\tilde{t}_R})  
 + \frac{m_t A_t}{m^2_{\tilde{q}}}P^{RL2}_{\gamma }(x_{\tilde{W}}, x_{\tilde{t}_R}) \bigg\} \bigg], \notag \\
 C_Z &= \frac{\alpha_e}{4\pi \lambda_t \sin^2 \theta_W}\bigg[ (\delta^u_{LL})_{32}
 \frac{m_t M_2}{m^2_{\tilde{q}}}\frac{m_t A_t}{m^2_{\tilde{q}}} P^{LL}_Z(x_{\tilde{W}}, x_{\mu }, x_{\tilde{t}_R}) 
 \notag\\&\quad
 + (\delta^u_{RL})_{32}\bigg\{ \frac{m_t M_2}{m^2_{\tilde{q}}}P^{RL1}_Z(x_{\tilde{W}}, x_{\mu }, x_{\tilde{t}_R}) 
 + \frac{m_t A_t}{m^2_{\tilde{q}}}P^{RL2}_Z(x_{\tilde{W}}, x_{\tilde{t}_R}) \bigg\} \bigg],\notag \\
 C_{B_u} &= \frac{\alpha_e}{4\pi \lambda_t \sin^2 \theta_W} \frac{M_W^2}{m^2_{\tilde{q}}}\bigg[ (\delta^u_{LL})_{32} 
 \bigg\{ \frac{M_2^2}{m^2_{\tilde{q}}}B^{LL1}_u (x_{\tilde{W}}) 
 + \frac{m_t M_2}{m^2_{\tilde{q}}}\frac{m_t A_t}{m^2_{\tilde{q}}}B^{LL2}_u (x_{\tilde{W}}, x_{\mu }, x_{\tilde{t}_R}) \bigg\} 
 \notag\\&\quad
 + (\delta^u_{RL})_{32}\bigg\{ \frac{m_t M_2}{m^2_{\tilde{q}}}B^{RL1}_u (x_{\tilde{W}}, x_{\mu }, x_{\tilde{t}_R})
 + \frac{M_2^2}{m^2_{\tilde{q}}}\frac{m_t A_t}{m^2_{\tilde{q}}}B^{RL2}_u (x_{\tilde{W}}, x_{\tilde{t}_R}) \bigg\}   \bigg],\notag \\
 C_{B_d} &= \frac{\alpha_e}{4\pi \lambda_t \sin^2 \theta_W} \frac{M_W^2}{m^2_{\tilde{q}}}\bigg[ (\delta^u_{LL})_{32} 
 \bigg\{ B^{LL1}_d (x_{\tilde{W}}) 
 + \frac{m_t M_2}{m^2_{\tilde{q}}}\frac{m_t A_t}{m^2_{\tilde{q}}}B^{LL2}_d (x_{\tilde{W}}, x_{\mu }, x_{\tilde{t}_R}) \bigg\}
 \notag\\&\quad
 + (\delta^u_{RL})_{32}\bigg\{ \frac{m_t M_2}{m^2_{\tilde{q}}}B^{RL1}_d (x_{\tilde{W}}, x_{\mu }, x_{\tilde{t}_R})
 + \frac{m_t A_t}{m^2_{\tilde{q}}}B^{RL2}_d (x_{\tilde{W}}, x_{\tilde{t}_R}) \bigg\}   \bigg],
 \label{eq:chargino}
\end{align}
where the trilinear coupling of the scalar top, $A_t$, is defined as $\mathcal{L} = -m_t A_t \tilde t_R^* \tilde t_L+ {\rm h.c.}$. The contributions to $C_8$ and $C_{10}$ are suppressed. The chargino contributions to the magnetic dipole operators are obtained as
\begin{align}
 C_{7\gamma } &= \frac{1}{\lambda_t}\frac{M_W^2}{m^2_{\tilde{q}}}
 \bigg[ (\delta^u_{LL})_{32} \bigg\{ \frac{M_2 \mu_H \tan \beta }{m^2_{\tilde{q}}}D^{LL1}_E( x_{\tilde{W}},x_{\mu }) + D^{LL2}_E(x_{\tilde{W}})  \bigg\} \notag \\
 &\quad
 +  (\delta^u_{RL})_{32} \bigg\{  \frac{m_t M_2}{m^2_{\tilde{q}}}
 D^{RL1}_E(x_{\tilde{W}},x_{\mu },x_{\tilde{t}_R})  \notag \\
 &\quad
 +\frac{M_2 \mu_H \tan \beta }{m^2_{\tilde{q}}} \frac{m_t A_t}{m^2_{\tilde{q}}}
 D^{RL2}_E(x_{\tilde{W}},x_{\mu },x_{\tilde{t}_R}) 
 +\frac{m_t A_t}{m^2_{\tilde{q}}}D^{RL3}_E(x_{\tilde{W}}, x_{\tilde{t}_R})  \bigg], \notag \\
  C_{8G} &= \frac{1}{\lambda_t}\frac{M_W^2}{m^2_{\tilde{q}}}
 \bigg[ (\delta^u_{LL})_{32} \bigg\{ \frac{M_2 \mu_H \tan \beta }{m^2_{\tilde{q}}}D^{LL1}_C( x_{\tilde{W}},x_{\mu }) + D^{LL2}_C(x_{\tilde{W}})  \bigg\} \notag \\
 &\quad
 +  (\delta^u_{RL})_{32} \bigg\{  \frac{m_t M_2}{m^2_{\tilde{q}}}
 D^{RL1}_C(x_{\tilde{W}},x_{\mu },x_{\tilde{t}_R})  \notag \\
 &\quad
 +\frac{M_2 \mu_H \tan \beta }{m^2_{\tilde{q}}} \frac{m_t A_t}{m^2_{\tilde{q}}}
 D^{RL2}_C(x_{\tilde{W}},x_{\mu },x_{\tilde{t}_R}) 
 +  \frac{m_t A_t}{m^2_{\tilde{q}}}D^{RL3}_C(x_{\tilde{W}}, x_{\tilde{t}_R}) \bigg].
 \label{eq:chargino-magnetic}
\end{align}
All the loop functions are summarized in App.~\ref{app:functions}, and the parameters are defined as
\begin{equation}
 x_{\tilde{W}} = \left( \frac{M_2}{m_{\tilde{q}}} \right)^2,\ \ \ x_{\mu } = \left( \frac{\mu_H}{m_{\tilde{q}}} \right)^2,\ \ \ 
 x_{\tilde{t}_R} = \left( \frac{m_{\tilde{t}_R}}{m_{\tilde{q}}} \right)^2,\notag \ \ \ 
\end{equation}
where $M_2$ and $\mu_H$ are the Wino and the Higgsino mass parameters, respectively. In the chargino contributions, the right-handed stop mass, $m_{\tilde{t}_R}$, is left explicit, because the Z-penguin contribution becomes larger when it is hierarchically small. Since the strange Yukawa coupling is neglected, the right-handed currents do not receive corrections. 

In the superCKM basis, the CKM matrix appears in the chargino--quark--squark vertices, which is another source of the flavor violation. However, this effect is discarded in this paper, because we are interested in the cases when the corrections are dominated by the squark mixings. 

There are no chargino contributions from $(\delta^u_{RR})_{32}$ and $(\delta^u_{LR})_{32}$, when the Yukawa coupling constants of the second generation are neglected. This is because the Wino couples only to the left-handed quarks. When the charm Yukawa coupling is turned on, the contributions become finite. The leading contributions are obtained as
\begin{align}
 C_Z &= \frac{\alpha_e}{4\pi \lambda_t \sin^2 \theta_W} \bigg[ 
 (\delta^u_{RR})_{32} \frac{m_c M_2}{m^2_{\tilde{q}}}\frac{m_t A_t}{m^2_{\tilde{q}}} P^{RR}_Z(x_{\tilde{W}}, x_{\mu }, x_{\tilde{t}_R}) \notag \\
 &\quad
 + (\delta^u_{LR})_{32} \bigg\{ \frac{m_c M_2}{m^2_{\tilde{q}}}P^{LR1}_Z(x_{\tilde{W}}, x_{\mu })
 + \frac{m_c m_t}{2M_W^2 \sin^2\beta }\frac{m_t A_t}{m^2_{\tilde{q}}}P^{LR2}_Z(x_{\mu }, x_{\tilde{t}_R})  \bigg\} \bigg],\notag \\
 C_{7\gamma} &= \frac{1}{\lambda_t}\frac{m_c \mu_H \tan \beta }{m^2_{\tilde{q}}} 
 \bigg[  (\delta^u_{RR})_{32} \frac{m_t A_t}{m^2_{\tilde{q}}}D^{RR}_E(x_{\mu }, x_{\tilde{t}_R}) 
 +  (\delta^u_{LR})_{32}D^{LR}_E(x_{\mu })  \bigg], \notag \\
  C_{8G} &= \frac{1}{\lambda_t}\frac{m_c \mu_H \tan \beta }{m^2_{\tilde{q}}} 
 \bigg[  (\delta^u_{RR})_{32} \frac{m_t A_t}{m^2_{\tilde{q}}}D^{RR}_C(x_{\mu }, x_{\tilde{t}_R}) 
 +  (\delta^u_{LR})_{32}D^{LR}_C(x_{\mu })  \bigg],
 \label{eq:chargino-right}
\end{align}
where the loop functions are in App.~\ref{app:functions}. The other contributions are negligible. Although they are suppressed by the charm Yukawa coupling, since the constraints become weaker, the mixings can be sizable. It is found that the contributions to $C_{7\gamma}$ and $C_{8G}$ are not suppressed by $M_W^2/m^2_{\tilde{q}}$, but rather, they include $\tan\beta$. 

The Wilson coefficients are set at the scale where the heavy fields decouple, i.e., the weak scale for the SM and the SUSY scale for the SUSY contributions. They evolve from the input scale down to the $m_b$ one by solving the renormalization group equations (see, e.g., \cite{Buchalla:1995vs}). Then, the $b-s$ transition amplitudes are evaluated. In the next section, the gluino and chargino contributions are studied on the CP asymmetries of the $B \to K\pi$ decays in the MI approximation.

Before proceeding, let us compare the MI formula to the full evaluation of the loop diagrams. The full method means that the loop diagrams are evaluated in the mass eigenstate basis. The calculations are easy to preform, whereas it is relatively hard to understand the underlying physics. This is because the full formula includes cancellations, e.g., by the GIM mechanism. In contrast, the MI formula is suited for understanding the process, because the mass matrices are expanded, and such cancellations are absent. Since the calculations are involved because there are a lot of diagrams after the mass matrices are expanded, we compared the results of the MI formula to those of the full method in order to confirm them. The full evaluations are found in the literature (see \cite{Hofer:2010ee} as a recent one). We checked that there is a good agreement between the two approaches in most of the parameter space. Exceptionally, when the right-handed stop is much lighter than the other superparticles, a subleading contribution can be dominant, because a corresponding loop function is enhanced. Then, the next-to-leading order evaluation is required. Otherwise, the MI approximation works well. 

In the MI approximation, a part of the chargino contributions has been missed in the literature \cite{Chakraverty:2003uv}. In particular, the contribution with $A_t$ has not been found, though those to $C_{7\gamma}$ and $C_{8G}$ are sizable. In the SUSY models, $A_t$ is considered to be comparable to the gluino mass, for instance, by the renormalization group evolution. Further, the recent LHC searches for the Higgs boson \cite{Higgs} may indicate relatively large $A_t$ \cite{Okada:1990gg}. Therefore, the contributions could be crucial for studies on the FCNCs. 
Moreover, the chargino mass matrix has not been expanded in \cite{Chakraverty:2003uv}. Since the mixing between the gaugino and the Higgsino stems from the electroweak symmetry breaking, the off-diagonal components of the mass matrix can be expanded, which is considered in the present result. This enables us to avoid the cancellations which appear in \cite{Chakraverty:2003uv}.


\section{$B \to K \pi$}
\label{sec:btokpi}

The difference of the CP asymmetries of the $B \to K\pi$ decays has been measured as \cite{Asner:2010qj}
\begin{align}
 \Delta A_{\rm CP} = A_{\rm CP}(B^+ \to K^+ \pi^0) - A_{\rm CP}(B^0 \to K^+ \pi^-) = 0.124 \pm 0.022,
 \label{eq:Kpiexp}
\end{align}
where the errors are added in quadrature. This is unexpected in the SM. The amplitude of the each decay mode can be decomposed by the topology of the weak transitions \cite{Gronau:1994rj},
\begin{align}
 \label{eq:decomposition}
 -\mathcal{A}(B^0 \to K^- \pi^+) &\simeq P + \frac{2}{3} P_{\rm EW}^C + T \\
 &= P \left[1 + \frac{2}{3} r_{\rm EW}^C e^{i\delta_{\rm EW}^C} + r_T e^{i\delta_T} e^{-i\phi_3} \right], \notag \\
 -\sqrt{2} \mathcal{A}(B^- \to K^- \pi^0) &\simeq P + P_{\rm EW} + \frac{2}{3} P_{\rm EW}^C + T + C \notag \\
 &= P \left[1 + r_{\rm EW} e^{i\delta_{\rm EW}}+ \frac{2}{3} r_{\rm EW}^C e^{i\delta_{\rm EW}^C} + 
 \left( r_T e^{i\delta_T} + r_C e^{i\delta_C} \right)e^{-i\phi_3} \right], \notag 
\end{align}
where the QCD penguin $P$, the color-allowed EW penguin $P_{\rm EW}$, and the color-suppressed EW penguin
$P_{\rm EW}^C$ are approximately proportional to $\lambda_c$ in the SM, while the color-allowed tree $T$ and the color-suppressed tree $C$ are proportional to $\lambda_u$. Hence, a relative weak phase, $\lambda_u/\lambda_c \propto e^{-i\phi_3}$, arises at the second line in the each amplitude. The strong phase relative to $P$ is denoted by $\delta_i$. The ratios of the magnitudes are defined by
\begin{align}
 r_{\rm EW} = |P_{\rm EW}/P|,~
 r_{\rm EW}^C = |P_{\rm EW}^C/P|,~
 r_T = |P_T/P|,~
 r_C = |P_C/P|.
\end{align}
In the decay amplitudes, the hadronic matrix elements are evaluated by using the QCD factorization as \cite{Beneke:2001ev,Hofer:2010ee},
\begin{align}
 P &\simeq \lambda_c ( a_4^c + r_\chi^K a_6^c ) A_{\pi K},~~~
 T \simeq \lambda_u a_1 A_{\pi K},~~~
 C \simeq \lambda_u a_2 A_{K \pi}, \notag \\
 P_{\rm EW} &\simeq \frac{3}{2} \lambda_c \left( - a_7 + a_9 \right) A_{K \pi},~~~
 P_{\rm EW}^C \simeq \frac{3}{2} \lambda_c \left( a_{10}^c + r_\chi^K a_8^c \right) A_{\pi K},
 \label{eq:QCDF}
\end{align}
where subleading contributions including the weak annihilation are omitted for simplicity. The definitions of the parameters are found in \cite{Beneke:2001ev}. The SM naively satisfies $1 > r_T, r_{\rm EW} > r_{\rm EW}^C, r_C$ (see, e.g., \cite{Hofer:2010ee}). The difference of the CP asymmetries is represented as 
\begin{align}
 \Delta A_{\rm CP} \simeq -2 r_C \sin \delta_C \sin \phi_3 
 + 2 r_{\rm EW}\, r_T \sin ( \delta_{\rm EW} + \delta_T ) \sin \phi_3.
 \label{eq:DeltaAcp}
\end{align}
Since the SM predicts $r_T, r_{\rm EW} = O(0.1)$ and $r_{\rm EW}^C, r_C = O(0.01)$, $\Delta A_{\rm CP} \sim 0.01$ is expected, which is smaller than the experimental result, (\ref{eq:Kpiexp}). 
 
If the NP is responsible for the discrepancy, the SM evaluation (\ref{eq:DeltaAcp}) is affected. The NP contributions are  parameterized as
\begin{align}
 P + \frac{2}{3} P_{\rm EW}^C + T &\to P + \frac{2}{3} P_{\rm EW}^C + T + X_{\rm NP}, \notag \\
 P_{\rm EW} + C &\to P_{\rm EW} + C + Y_{\rm NP},
 \label{eq:NP}
\end{align}
where $X_{\rm NP}$ represents the NP contribution common to $B^0 \to K^- \pi^+$ and $B^- \to K^- \pi^0$, while $Y_{\rm NP}$ contributes only to $B^- \to K^- \pi^0$. Since the NP generally involves new CP-violating phases, there are four types of the corrections to $\Delta A_{\rm CP}$:
\begin{enumerate}
\item \label{enumA}
When $X_{\rm NP}$ is dominant with a sizable weak phase relative to $P({\rm SM})$, the isospin violation is from $P_{\rm EW} = P_{\rm EW}({\rm SM})$, and $\Delta A_{\rm CP}$ is evaluated as
\begin{align}
 \Delta A_{\rm CP}({\rm NP}) \simeq 2 r_{\rm EW}\, r_X \sin ( \delta_{\rm EW} + \delta_X ) \sin \phi_X,
 \label{eq:enumA}
\end{align}
where $r_X, \delta_X$ and $\phi_X$ are $|X_{\rm NP}/P({\rm SM})|$, the strong and the weak phases relative to that of $P({\rm SM})$, respectively. Thus, $X_{\rm NP} \sim P({\rm SM})$ is required to account for (\ref{eq:Kpiexp}).
\item \label{enumB}
When $X_{\rm NP}$ is large, but its weak phase is almost aligned to $P({\rm SM})$, $X_{\rm NP}$ contributes to $P$ as $P \to P + X_{\rm NP}$. Then, $\Delta A_{\rm CP}$ has the same form as (\ref{eq:DeltaAcp}). In order to enhance it, $r_i$ in (\ref{eq:DeltaAcp}) needs to be magnified. Thus, $X_{\rm NP} \simeq -P({\rm SM})$ is required. Note that the branching ratios of $B \to K\pi$ are also suppressed.
\item \label{enumC}
If $Y_{\rm NP}$ dominates the NP contributions with a large weak phase, $\Delta A_{\rm CP}$ becomes 
\begin{align}
 \Delta A_{\rm CP}({\rm NP}) \simeq -2 r_Y \sin \delta_Y \sin \phi_Y,
 \label{eq:AcpC}
\end{align}
where the parameters are defined similarly to (\ref{eq:enumA}). Then, $|Y_{\rm NP}|$ is required to be at least comparable to $|P_{\rm EW}({\rm SM})|$ for (\ref{eq:Kpiexp}).
\item \label{enumD}
If $Y_{\rm NP}$ is large, but the weak phase is suppressed, $Y_{\rm NP}$ contributes to $P_{\rm EW}$. Noting that the CP violation originates in $T({\rm SM})$, $\Delta A_{\rm CP}$ becomes
\begin{align}
 \Delta A_{\rm CP}({\rm NP}) \simeq 2 r_Y\, r_T \sin ( \delta_Y + \delta_T ) \sin \phi_3.
\end{align}
The experimental result (\ref{eq:Kpiexp}) indicates that $Y_{\rm NP}$ is as large as $P({\rm SM})$.
\end{enumerate}
It is commented that when the NP appears in the right-handed currents, their contributions are evaluated by extending $C_i$ in \cite{Beneke:2001ev} as $C_i \to C_i - \tilde C_i$, where the relative sign is determined by the parity of the final state \cite{Kagan:2004ia}. Although the SUSY is unlikely to contribute to the strong phases \cite{Datta:2004re}, since there are discussions \cite{Hofer:2010ee}, we focus on the magnitude of the SUSY contributions in the following study. If none of the conditions, \ref{enumA}--\ref{enumD}, is satisfied, the measured difference (\ref{eq:Kpiexp}) is not explained by SUSY, irrespective of the strong phase. 

In many cases, the SUSY contributions are dominated by the chromomagnetic dipole term, $C_{8G}$. According to the MI formula, it is larger than the four-quark operators. From the relation between $a_i$ and $C_i$ \cite{Beneke:2001ev}, $P$ in (\ref{eq:QCDF}) receives a correction as
\begin{align}
 a_4 + r_\chi^K a_6 = - \frac{C_F \alpha_s}{2\pi N_c} (I_K + r_\chi^K) C_{8G}^{\rm eff} + \ldots,
 \label{eq:Kpi8g}
\end{align}
where $I_K$ is $\int_0^1 dx\, \Phi_K / (1-x)$, and $\Phi_K$ is the leading-twist distribution amplitude of the K meson. It is emphasized that this contribution corresponds to $X_{\rm NP}$, whose weak phase is from the squark mixings in $C_{8G}$. 

The SUSY contributions to $C_{8G}$ is tightly constrained by ${\rm Br}(b \to s\gamma)$. This is because the structure of $C_{7\gamma}$ is very similar to $C_{8G}$, as shown in Sec.~\ref{sec:wilson} explicitly. Since the experimental results \cite{Asner:2010qj} agree well with the SM prediction \cite{Misiak:2006zs}, extra corrections are restricted to be within the range,
\begin{align}
 -0.28 \times 10^{-4} < \Delta {\rm{B}}(b \to s\gamma) < 1.08 \times 10^{-4},
 \label{eq:delBrbsg}
\end{align}
at the $2\sigma$ level. This means that $C_{7\gamma}$ is around 0.3 at the $m_b$ scale. In addition, the recent measurements of $B \to K^*\mu^+\mu^-$ support that $C_{7\gamma} C_9^{\ell}$ is consistent with the SM, where $C_9^{\ell}$ is the Wilson coefficient of the effective operator, $\mathcal{O}_9^{\ell} = (\bar s_L \gamma_\mu b_L)(\bar\ell \gamma^\mu \ell)$ \cite{Beaujean:2012uj}. Since the sign of $C_9^{\ell}$ is not flipped by the SUSY contributions \cite{Ali:2002jg}, $C_{7\gamma}$ is dominated by the SM\footnote{
The amplitude also receives contributions from the chargino and charged Higgs diagrams with the flavor violation by the CKM matrix. They are assumed to be cancelled to each other for simplicity. Even if they are included, the following conclusion does not change qualitatively. 
}.

\begin{figure}[t]
\begin{center}
\includegraphics[scale=0.7]{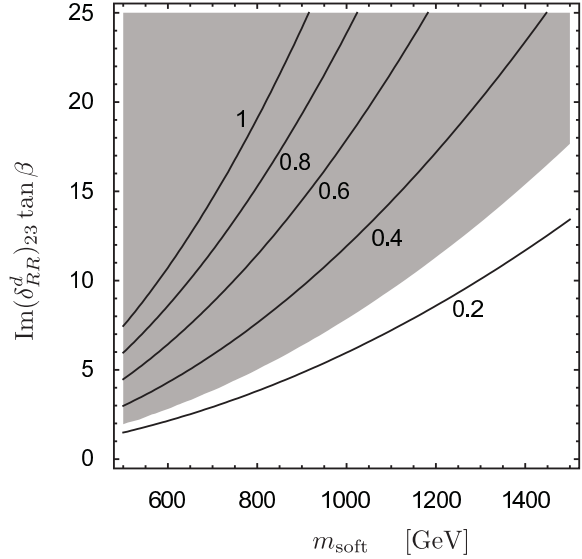}
\caption{Contours of $|X_{\rm NP}/P({\rm SM})|$ are shown for 0.2, 0.4, 0.6, 0.8 and 1. The gray shaded region is excluded by ${\rm Br}(b \to s\gamma)$ at the $2\sigma$ level. The parameter $m_{\rm soft}$ represents the soft breaking parameters and $\mu_H$. The squark mixing other than ${\rm Im}\,(\delta^d_{RR})_{23}$ are set to be zero. }
\label{fig:penguin}
\end{center}
\end{figure}

In practice, when $(\delta^d_{LL})_{23}$ and $(\delta^d_{RR})_{23}$ are considered, $C_{7\gamma}$ and $C_{8G}$ are larger than the four-quark operators by $\tan\beta$ (see (\ref{eq:gluino})). They contribute to $X_{\rm NP}$ and $b \to s\gamma$. In Fig.~\ref{fig:penguin}, $|X_{\rm NP}/P({\rm SM})|$ and the $b \to s\gamma$ bound (\ref{eq:delBrbsg}) are shown for $m_{\rm soft}$ and ${\rm Im}(\delta^d_{RR})_{23} \times \tan\beta$. It is found that $X_{\rm NP}$ is less than 30\% of $P({\rm SM})$ under the constraint, and the maximum is insensitive to $m_{\rm soft}$. On the other hand, $Y_{\rm NP}$ is found to be tiny. If the photon penguins are compared to $C_{8G}$, (\ref{eq:gluino-photon}) is proportional to $\alpha_e$ without $\tan\beta$. Although the Z penguins can be comparable to the photon penguin, they do not dominate the EW penguins at least for $\tan\beta \lesssim 50$, because they are proportional to $m_b$ (see (\ref{eq:gluino-Z})). Hence, $Y_{\rm NP}$ is estimated as
\begin{align}
 Y_{\rm NP} \simeq P_{\rm EW}({\rm SUSY}) 
 \sim \frac{2\pi N_c}{C_F \alpha_s (I_K + r_\chi^K)} \frac{C_{7,9}}{C_{8G}}\, P({\rm SUSY})
 \sim \frac{\alpha_e}{\tan\beta} P({\rm SUSY}),
 \label{eq:YPratio}
\end{align}
where the loop functions are taken into account in the last relation. Since $r_{\rm EW}({\rm SM})$ is of order $0.1$, $Y_{\rm NP}$ is found to be less than percent of $P_{\rm EW}({\rm SM})$. Therefore, none of the conditions, \ref{enumA}--\ref{enumD}, is satisfied, and the measurement (\ref{eq:Kpiexp}) is not explained, because of the tight constraint by $b \to s\gamma$. 

The situation is very similar for $(\delta^d_{LR})_{23}$ and $(\delta^d_{RL})_{23}$. The magnetic dipole terms in (\ref{eq:gluino}) are enhanced by $m_{\tilde g}/m_b$, whereas $Y_{\rm NP}$ from (\ref{eq:gluino-Z}) is proportional to $m_b \tan\beta$ without $1/(G_F m_{\tilde q}^2)$. Then, $C_{7,9}/C_{8G}$ in (\ref{eq:YPratio}) is estimated to be $\sim G_F m_b^2 \tan\beta$, which leads to $Y_{\rm NP}/P_{\rm EW}({\rm SM}) \lesssim 1\%$ under the constraint from $b \to s\gamma$. This is too small to explain (\ref{eq:Kpiexp}).

In the case of the chargino contributions with $(\delta^u_{LL})_{32}$, the magnetic dipole terms in (\ref{eq:chargino-magnetic}) include $\tan\beta$, though they are proportional to $M_W^2$. As seen from (\ref{eq:chargino}), the Z penguin is not suppressed by $M_W^2$, whereas it is proportional to $\alpha_e m_t^2$. Thus, $C_{7,9}/C_{8G} \sim \alpha_e m_t^2/(M_W^2 \tan\beta)$ is obtained for (\ref{eq:YPratio}). The SUSY contributions are controlled by $C_{7\gamma}$ and $C_{8G}$, and (\ref{eq:Kpiexp}) is not explained by $(\delta^u_{LL})_{32}$ by $b \to s\gamma$. 

\begin{figure}[t]
\begin{center}
\includegraphics[scale=0.7]{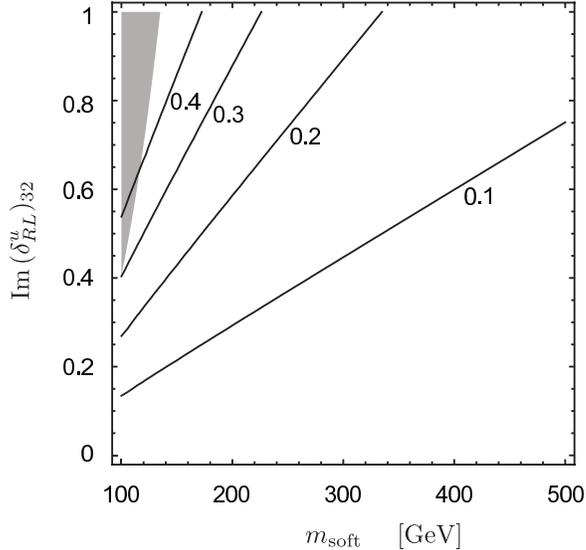}
\caption{Contours of $|Y_{\rm NP}/P_{\rm ew}({\rm SM})|$ are shown for 0.1, 0.2, 0.3 and 0.4. It does not reach 0.5. The gray shaded region is excluded by ${\rm Br}(b \to s\gamma)$ at the $2\sigma$ level. The parameter $m_{\rm soft}$ represents the soft breaking parameters and $\mu_H$, while $A_t = 0$ is taken. The squark mixing other than ${\rm Im}\,(\delta^u_{RL})_{32}$ are set to be zero. }
\label{fig:ewpenguin}
\end{center}
\end{figure}

The $b \to s\gamma$ constraint is relaxed, when $(\delta^u_{RL})_{32}$ is considered and if $A_t$ is suppressed, as found from (\ref{eq:chargino}) and (\ref{eq:chargino-magnetic}). There is no enhancement for $C_{8G}$ compared to the gluon penguin, $C_g$, in the SUSY contributions to the decay amplitudes. Since the Z penguin is not proportional to $M_W^2$, the ratio in (\ref{eq:YPratio}) becomes $C_{7,9}/C_{8G} \sim \alpha_e m_{\tilde q}^2/M_W^2$. Thus, $Y_{\rm NP}$ can be larger than $X_{\rm NP} \sim P({\rm SUSY})$ for $m_{\tilde q} \sim 1$TeV, though it decreases as $m_{\tilde q}$ is lowered. In Fig.~\ref{fig:ewpenguin}, contours of $|Y_{\rm NP}/P_{\rm EW}({\rm SM})|$ and $b \to s\gamma$ are plotted for $m_{\rm soft}$ and ${\rm Im}(\delta^u_{RL})_{32}$. It is seen that $Y_{\rm NP}$ can reach about 40\% of $P_{\rm EW}({\rm SM})$, namely a few percents of $P({\rm SM})$, for $m_{\rm soft} \simeq 150$GeV. Then, $\Delta A_{\rm CP}({\rm SUSY})$ could be as large as $O(0.01)$ from (\ref{eq:AcpC}), if the $b \to s \gamma$ bound is imposed.

There are constraints on $(\delta^u_{RL})_{32}$ other than from $b \to s \gamma$ such as $B \to K^*\mu^+\mu^-$ \cite{Behring:2012mv} and the vacuum stability \cite{Park:2010wf}. According to the recent analysis based on the updated experimental data of $B \to K^*\mu^+\mu^-$ \cite{Behring:2012mv}, $(\delta^u_{RL})_{32}$ is limited to be less than $\sim 0.1$ for $m_{\rm soft} \sim 100$GeV and $\sim 0.3$ for $m_{\rm soft} \sim 1$TeV. The vacuum stability condition \cite{Park:2010wf} provides $(\delta^u_{RL})_{32} \lesssim 0.5$ and 0.3 for $m_{\rm soft} \sim 500$GeV and 1TeV, respectively. Thus, $Y_{\rm NP}$ is bounded to be less than 10\% of $P_{\rm EW}({\rm SM})$ according to Fig.~\ref{fig:ewpenguin}. Thus, $(\delta^u_{RL})_{32}$ cannot be large enough to explain (\ref{eq:Kpiexp}). Furthermore, since direct searches for the superparticles already excluded the light mass region \cite{LHCSUSY}, the bound is likely to be much severer, and it is concluded that $(\delta^u_{RL})_{32}$ cannot account for (\ref{eq:Kpiexp}).

In the above study on $(\delta^u_{RL})_{32}$, $A_t$ was assumed to vanish. As the coupling increases, the magnetic dipole operators are enhanced (see (\ref{eq:chargino-magnetic})). Then, the $b \to s\gamma$ bound becomes severer, and $Y_{\rm NP}$ is disallowed to be large. On the other hand, when the right-handed stop is much lighter than the other superparticles, the loop function of the Z penguin increases (see App.~\ref{app:functions}). However, the bound from $B \to K^*\mu^+\mu^-$ simultaneously becomes severer \cite{Behring:2012mv}. We could not find the region where the difference of the CP asymmetries of the $B \to K\pi$ decays (\ref{eq:Kpiexp}) is explained by $(\delta^u_{RL})_{32}$. 

Let us comment on the right-handed mixings of the up-type squarks, $(\delta^u_{RR})_{32}$ and $(\delta^u_{LR})_{32}$. As found in (\ref{eq:chargino-right}), their contributions are proportional to the charm mass, $m_c$. Since the contribution to $C_{7\gamma}$ and $C_{8G}$ are suppressed, the mixings can be as large as unity. However, it is estimated that $Y_{\rm NP}$ reaches less than 1\% of $P_{\rm EW}({\rm SM})$ even for $m_{\rm soft} = 100$GeV, $\tan\beta = 40$ and ${\rm Im}(\delta^u_{LR})_{32} = 1$. Thus, it is impossible to solve the $K\pi$ puzzle by $(\delta^u_{RR})_{32}$ and $(\delta^u_{LR})_{32}$.

In the QCD factorization, one of the largest uncertainties stems from parameterizations of the divergences in the soft gluon interaction and the weak annihilation \cite{Beneke:2001ev}. So far, the scenario S4 in \cite{Beneke:2001ev} was applied to the estimation of (\ref{eq:QCDF}). If the parameters are varied, $\Delta A_{\rm CP}$ can change by $\sim 5$\% (see, e.g., \cite{Hofer:2010ee}). Nonetheless, we checked that the SUSY contributions are irrelevant for the $K\pi$ puzzle.

The above study is compared to the previous works which investigated the chargino contributions for the $K\pi$ puzzle\cite{Khalil:2005qg,Khalil:2006hj,Khalil:2009zf,Hofer:2010ee}. The authors in \cite{Khalil:2005qg,Khalil:2006hj,Khalil:2009zf} insisted that the puzzle was solved, whereas those in \cite{Hofer:2010ee} obtained the opposite conclusion. The difference exists in the method to evaluate the loop diagrams: the former used the MI approximation, while the latter relied on the full method. In this paper, we revisited the $K\pi$ puzzle with the MI formula and obtained the same conclusion as \cite{Hofer:2010ee}. Although the MI formula used in \cite{Khalil:2005qg,Khalil:2006hj} missed some contributions which mentioned in the previous section, they may not be the leading reason of the contradiction. We found that some numerical results in \cite{Khalil:2005qg,Khalil:2006hj} are not reproduced, which could cause the difference. On the other hand, the strong phase could be a source of the gap between \cite{Khalil:2005qg,Khalil:2006hj,Khalil:2009zf} and \cite{Hofer:2010ee}, as mentioned in \cite{Hofer:2010ee}. In the former analysis, the phase is taken to be free, while it is fixed in the QCD factorization in the latter. In this paper, the puzzle was studied with the strong phase of the SUSY contributions supposed to be a free parameter. It was concluded that the chargino contributions are tightly restricted irrespective to the strong phase. Thus, it is believed that the SUSY is not responsible for the $K\pi$ puzzle. 

Let us comment on a loophole of the above study. The SUSY contributions could be cancelled to each other for $C_{7\gamma}$. Then, large squark mixings might be considered to explain the anomaly. In that case, careful analyses are required, because such large mixings easily spoil other experimental constraints. For instance, when $(\delta^d_{LL})_{23}$ is as large as unity with $(\delta^d_{LR})_{23}$ tuned to cancel $C_{7\gamma}$, SUSY contributions to $B_s \to J/\psi \phi$ as well as $\Delta m_s$ may conflict with the experimental results \cite{Buras:2012ts}. Anyway, the complete study is devoted for future works. 


\section{Conclusion}
\label{sec:conclusion}

The MI formula of $b \to s \bar q q$ was provided for the chargino contributions as well as those of the gluino. This approach has an advantage that they are easier to understand compared to the full evaluation. For instance, the cancellations which appear in the latter method such as the GIM mechanism are performed automatically, and it is possible to find out relevant  contributions. 

The MI formula was, then, applied to the analysis of the $K\pi$ puzzle. It was found that the SUSY models are not appropriate to explain the experimental result of the discrepancy between the CP asymmetries of $B^0 \to K^+ \pi^-$ and $B^\pm \to K^\pm \pi^0$. Since the structure of the SUSY contributions is clearly seen by means of the MI approximation, the conclusion is expected to be less dependent on details of the models. 

Let us comment on a future prospect of the $K\pi$ puzzle. The CP asymmetries of the $B \to K\pi$ decays follows a sum rule unless the anomaly is due to the EW penguin \cite{Gronau:2005kz}. If the violation of the sum rule will be observed in future, physics beyond the SM is indicated other than the SUSY models.

\section*{Acknowledgments}
This work was supported by Grand-in-Aid for Scientific research from
the Ministry of Education, Science, Sports, and Culture (MEXT), Japan,
No. 23740172 (M.E.).

\appendix

\section{Loop Functions}
\label{app:functions}

In this appendix, the loop functions in the gluino and chargino contributions are provided.

\subsection{Gluino contributions}

The loop functions that appear in (\ref{eq:gluino}), (\ref{eq:gluino-photon}) and (\ref{eq:gluino-Z}) are given by 
\begin{align}
 &B_1(x) = \frac{1+4x-5x^2+4x\ln x +2x^2\ln x}{8(1-x)^4} ,\notag \\
 &B_2(x) = \frac{5-4x-x^2+2\ln x+4x\ln x}{2(1-x)^4} , \notag \\
 &P_1(x) = \frac{1-6x+18x^2-10x^3-3x^4+12x^3 \ln x}{18(x-1)^5} , \notag \\
 &P_2(x) = \frac{7-18x+9x^2+2x^3+3 \ln x-9x^2 \ln x}{9(x-1)^5} , \notag \\
 &M_1(x) = 4B_1(x) , \notag \\
 &M_2(x) = -xB_2(x) ,\notag \\
 &M_3(x) = \frac{-1+9x+9x^2-17x^3+18x^2\ln x+6x^3\ln x}{12(x-1)^5} , \notag \\
 &M_4(x) = \frac{-1-9x+9x^2+x^3-6x-6x^2\ln x}{6(x-1)^5} , \notag \\
 &M_a(x) = -3M_4(x) , \notag \\
 &M_b(x) = -\frac{3-3x^2+(1+4x+x^2)\ln x}{(x-1)^5} ,\notag \\
 &Z_{a}(x) = \frac{2x^3+3x^2-6x+1-6x^2 \ln x}{6(1-x)^4},\notag \\
 &Z_{b}(x) = \frac{x^4-8x^3+8x-1+12x^2\log x}{12(1-x)^5} , \notag 
 \label{eq:gluloopfuncMIA}
\end{align}

\subsection{Chargino contributions}
\subsubsection{The functions $f^{(j)}_{i}$, $g^{(j)}_{i}$, $h_i$ }
\label{app:fgh}

The chargino contributions are represented by the functions, $f^{(j)}_{i}$, $g^{(j)}_{i}$, $h_i$:
\begin{align}
 &f^{(0)}_1(x, y) = \frac{1}{x-y} -\frac{x( \log x -\log y )}{(x-y)^2},~~~~f^{(0)}_1(x) =  f^{(0)}_1(x, 1),    \notag \\ 
 &f^{(0)}_2(x,y) = f^{(0)}_1(y,x),~~~~f^{(0)}_2(x) =  f^{(0)}_2(x, 1), \notag \\
 &f^{(0)}_3(x,y) =  \frac{x+y}{2y(x-y)^2} -\frac{x(\log x - \log y)}{(x-y)^3} ,~~~~f^{(0)}_3(x) =  f^{(0)}_3(x, 1), \notag \\
 &f^{(0)}_4(x,y) = f^{(0)}_3(y,x) ,~~~~f^{(0)}_4(x) =  f^{(0)}_4(x, 1), \notag  \\
 &f^{(0)}_5(x) = \frac{2}{(x-1)^2} +\frac{1+x}{(x-1)^3}\log x ,\notag \\
 &f^{(0)}_6(x) = -\frac{5+x}{2(x-1)^3} +\frac{1+2x}{(x-1)^4}\log x ,\notag \\
 &f^{(1)}_1(x,y) = \frac{3 x-y}{2 (x-y)^2}-\frac{x^2( \log x -\log y ) }{(x-y)^3} ,~~~~f^{(1)}_1(x) =  f^{(1)}_1(x, 1), \notag \\
 &f^{(1)}_2(x,y) =  f^{(1)}_1(y,x) ,~~~~f^{(1)}_2(x) =  f^{(1)}_2(x, 1),  \notag  \\
 &f^{(1)}_3(x,y) = -2yf^{(0)}_3(x,y),~~~~f^{(1)}_3(x) =  f^{(1)}_3(x, 1),  \notag \\
 &f^{(1)}_4(x,y) = \frac{2x^2+5xy-y^2}{6y(x-y)^3} -\frac{x^2(\log x -\log y )}{(x-y)^4} , ~~~~f^{(1)}_4(x) =  f^{(1)}_4(x, 1),  \notag \\
 &f^{(1)}_5(x,y) = f^{(1)}_4(y,x) ,~~~~f^{(1)}_5(x) =  f^{(1)}_5(x, 1),   \notag \\
 &f^{(1)}_6(x) = -\frac{1+5x}{2(x-1)^3} +\frac{x(2+x)}{(x-1)^4}\log x,  \notag \\
 &f^{(1)}_7(x) = - f^{(0)}_6(x),  \notag \\
 &f^{(1)}_8(x) = -\frac{1+10x+x^2}{3x(x-1)^5} +\frac{2(1+x)}{(x-1)^5}\log x, \notag \\
 &f^{(2)}_1(x,y) = \frac{11x^2-7xy+2y^2}{6(x-y)^3} - \frac{x^3(\log x -\log y)}{(x-y)^4} , ~~~~f^{(2)}_1(x) =  f^{(2)}_1(x, 1),  \notag \\
 &f^{(2)}_2(x,y) = f^{(2)}_1(y,x) ,~~~~f^{(2)}_2(x) =  f^{(2)}_2(x, 1),   \notag \\
 &f^{(2)}_3(x) = \frac{1-5x+13x^2+3x^3}{12(x-1)^4} -\frac{x^3}{(x-1)^5}\log x ,\notag \\
 &f^{(2)}_4(x) = \frac{-17-8x+x^2}{6(x-1)^4} +\frac{1+3x}{(x-1)^5}\log x .\notag 
 \end{align}
The definition of the functions $g^{(j)}_{i}$:
\begin{align}
 &g^{(0)}_1(x,y,z) =  -\frac{x\log x}{(x-y)(x-z)}
 +(x \leftrightarrow y) + (x \leftrightarrow z) , \notag \\
 &g^{(1)}_1(x,y,z) = -\frac{x^2\log x}{(x-1)(x-y)(x-z)} +(x \leftrightarrow y) 
 + (x \leftrightarrow z) ,\notag \\
 &g^{(0)}_1(x,y) = g^{(0)}_1(x,y,1) ,\notag \\
 &g^{(1)}_1(x,y) = g^{(1)}_1(x,x,y) , \notag \\
 &g^{(1)}_2(x,y) = g^{(1)}_1(x,y,1) . \notag 
\end{align}
The definition of the functions $h_i$:
\begin{align}
 &h_1(x,y) = f^{(0)}_1(x,y) - f^{(1)}_1(x,y) , ~~~~h_1(x) = h_1(x,1) ,\notag \\
 &h_2(x,y) = f^{(1)}_1(x,y) - f^{(2)}_1(x,y) ,~~~~h_2(x) = h_2(x,1) ,\notag \\
 &h_3(x,y) = f^{(1)}_2(x,y) - f^{(2)}_2(x,y) , ~~~~h_3(x) = h_3(x,1) ,\notag \\
 &h_4(x) = 2f^{(0)}_3(x) -3f^{(1)}_4(x) ,\notag \\
 &h_5(x) = 3f^{(1)}_4(x) - 4f^{(2)}_3(x) ,\notag \\
 &h_6(x) = f^{(1)}_7(x) - f^{(2)}_4(x) ,\notag \\
 &h_7(x) = x f^{(1)}_8(x) -\frac{2}{3} f^{(2)}_4(x) ,\notag \\
 &h_8(x,y) = x f^{(1)}_5(x,y) -\frac{2}{3} f^{(2)}_2(x,y) , ~~~~h_8(x) = h_8(x,1) .\notag 
 \end{align}

\subsubsection{Penguin}
\label{app:pen}

The loop functions for $C_{g}$, $C_{\gamma }$ and $C_{Z}$ in (\ref{eq:chargino}): 
\begin{align}
 &P^{LL}_{g}(x) = -\frac{9}{2} f^{(2)}_3(x)  , \notag \\
 &P^{RL1}_{g}(x,y,z) =  \frac{9}{8}\frac{ f_1^{(2)}(x) - f_1^{(2)}(x,z) 
 - (x \leftrightarrow  y) }{(x-y)(1-z)} , \notag \\ 
 &P^{RL2}_{g} (x,y)= \frac{9}{8(1-y)}\bigg\{ 4f^{(2)}_3(x) 
 -\frac{f^{(2)}_1(x) -f^{(2)}_1(x,y) }{1-y} \bigg\}  ,\notag \\
 &P^{LL}_{\gamma }(x) = -4 f^{(2)}_3(x)  
  - \frac{9}{2} h_7(x) , \notag \\
 &P^{RL1}_{\gamma }(x,y,z) =  \frac{f_1^{(2)}(x) - f_1^{(2)}(x,z) 
 - (x \leftrightarrow  y) }{(x-y)(1-z)} \notag \\
 &~~~~~~~~~~~~~~~~~-\frac{9}{2(x-y)}\bigg\{ \frac{f^{(1)}_2(x) - f^{(1)}_2(x,z)}{1-z}  
 - \frac{g^{(1)}_1(x,z) - g^{(1)}_1(x,y,z)}{x-y} \bigg\}   , \notag \\ 
 &P^{RL2}_{\gamma } (x,y) = \frac{1}{1-y}\bigg\{ 4f^{(2)}_3(x) 
 -\frac{f^{(2)}_1(x) -f^{(2)}_1(x,y)}{1-y} \bigg\} \notag \\
  &~~~~~~~~~~~~~~~~~+\frac{9}{2(1-y)}\bigg\{ h_7(x) -\frac{ h_8(x) - h_8(x,y) }{1-y} \bigg\}  , \notag \\
  &P^{LL}_Z(x,y,z) =  \frac{f^{(1)}_1(x) -g^{(1)}_2(x,z)
 - (x \leftrightarrow y)}{(x-y)(1-z)}   \notag \\
 &~~~~~~~~~~~~~~~~~~+ \frac{f^{(1)}_3(x) - g^{(1)}_1(x,z)}{2(x-y)(1-z)}
  -\frac{g^{(1)}_2(x,z) - g^{(1)}_2(y,z)}{2(x-y)^2},\notag \\
 &P^{RL1}_Z(x,y,z) = -\frac{g^{(1)}_2(x,z) - g^{(1)}_2(y,z)  + g^{(1)}_1(x,z) - g^{(1)}_1(x,y,z)}
 {2(x-y)} ,\notag \\
 &P^{RL2}_Z(x,y) = -\frac{1}{1-y}\bigg\{ f^{(1)}_1(x) - g^{(1)}_2(x,y) 
 + xf^{(0)}_5(x) -\frac{1}{2}f^{(1)}_3(x)  \notag \\
 &~~~~~~~~~~~~~~~~~-\frac{x\big( f^{(0)}_2(x) - f^{(0)}_2(x,z) \big) }{1-y} 
 + \frac{1}{2}g^{(1)}_1(x,y) \bigg\}  .\notag 
\end{align}
The loop functions for $C_{Z}$ in (\ref{eq:chargino-right}):
\begin{align}
 &P^{RR}_Z (x,y,z) =  \frac{f^{(1)}_1(x) -g^{(1)}_2(x,z)
 - (x \leftrightarrow y)}{2(x-y)(1-z)}   \notag \\
 &~~~~~~~~~~~~~~~~~~+ \frac{f^{(1)}_3(x) - g^{(1)}_1(x,z)}{2(x-y)(1-z)}
  -\frac{g^{(1)}_2(x,z) - g^{(1)}_2(y,z)}{2(x-y)^2},\notag \\
 &P^{LR1}_Z(x,y) = P^{RL1}_Z(x,y,1) ,\notag \\
 &P^{LR2}_Z(x,y) = \frac{f^{(1)}_1(x) - f^{(1)}_1(y)}{x-y} . \notag 
\end{align}

\subsubsection{Box}
\label{app:box}

The loop functions for $C_{B_u}$ and $C_{B_d}$ in (\ref{eq:chargino}): 
\begin{align}
 &B^{LL1}_{u}(x) = \frac{2}{3}f^{(0)}_6(x)   ,\notag  \\
 &B^{LL2}_{u}(x,y,z) = \frac{2}{3(x-y)(1-z)}\bigg\{  f^{(1)}_6(x)
 -\frac{f^{(1)}_3(x) - g^{(1)}_1(x,z)}{1-z}  \notag \\
 &~~~~~~~~~~~~~~~~~~~~-\frac{f^{(1)}_1(x) - g^{(1)}_2(x,z) - (x \leftrightarrow y )}{x-y} \bigg\}   ,\notag  \\
 &B^{RL1}_{u}(x,y,z) =  -\frac{2}{3(x-y)}\bigg\{ \frac{f^{(1)}_3(x) - g^{(1)}_1(x,z) }{1-z}
 - \frac{ g^{(1)}_2(x,z) -  g^{(1)}_2(y,z)}{x-y} \bigg\}  ,\notag  \\
 &B^{RL2}_{u}(x,y) = - \frac{2}{3(1-y)}\bigg\{ f^{(0)}_6(x)  
 - \frac{1}{1-y}\bigg( f^{(0)}_5(x) -\frac{f^{(0)}_2(x) -f^{(0)}_2(x,y)}{1-y} \bigg) \bigg\}  ,\notag \\
 &B^{LL1}_{d}(x) = \frac{1}{3}f^{(1)}_6(x)   ,\notag  \\
 &B^{LL2}_{d}(x,y,z) = \frac{1}{2}B^{LL2}_{u}(x,y,z), \notag \\
 &B^{RL1}_{d}(x,y,z) = \frac{1}{2} B^{RL1}_{u}(x,y,z)  ,\notag  \\
 &B^{RL2}_{d}(x,y) = -\frac{1}{3(1-y)}\bigg\{ f^{(1)}_6(x) 
 -\frac{f^{(1)}_3(x) - g^{(1)}_1(x,y)}{1-y} \bigg\} . \notag 
\end{align}

\subsubsection{Magnetic and chromo-magnetic coefficients}
\label{app:mag}
The loop functions for $C_{7\gamma }$ and $C_{8G}$ in (\ref{eq:chargino-magnetic}):
\begin{align}
 &D^{LL1}_E(x,y) = \frac{4}{x-y}\big(  f_7(x) - f_7(y) \big)  ,  \notag \\
 &D^{LL2}_E(x) = -f^{(1)}_4(x) + \frac{10}{9}f^{(2)}_3(x)
  +\frac{  h_6(x) }{2}   ,\notag \\
 &D^{RL1}_E(x,y,z) = \frac{h_2(x) - h_2(x,z)  - (x \leftrightarrow y )  }{3(x-y)(1-z)} 
 +\frac{h_3(x) -h_3(x,z)  - (x \leftrightarrow y )  }{2(x-y)(1-z)} ,\notag \\
 &D^{RL2}_E(x,y,z) = -\frac{2}{3(x-y)(1-z)}\bigg\{ h_4(x)
 - \frac{h_1(x) -h_1(x,z) }{1-z} 
 - (x \leftrightarrow y) \bigg\}  \notag \\
 &~~~~~~~~~~~~~~~~~~~~-\frac{1}{(x-y)(1-z)}
 \bigg\{ f^{(1)}_7(x) -\frac{f^{(1)}_2(x) - f^{(1)}_2(x,z)}{1-z} 
 - (x \leftrightarrow y )\bigg\} ,\notag \\
 &D^{RL3}_E(x,y) = \frac{1}{3(1-y)}\bigg\{ h_5(x)
 -\frac{h_2(x) - h_2(x,y) }{1-y} \bigg\} \notag \\
 &~~~~~~~~~~~~~~~~~+\frac{1}{2(1-y)}\bigg\{  h_6(x)
 -\frac{h_3(x) - h_3(x,y) }{1-y} \bigg\} ,\notag \\
 &D^{LL1}_C(x,y) = \frac{4}{x-y} \big( f_8(x) - f_8(y) \big) ,\notag \\
 &D^{LL2}_C(x,y) = -\frac{3}{2}f^{(1)}_4(x) + \frac{5}{3}f^{(2)}_3(x) ,\notag \\
 &D^{RL1}_C(x,y,z) = \frac{h_2(x) - h_2(x,z)  - (x \leftrightarrow y )  }{2(x-y)(1-z)},\notag \\
 &D^{RL2}_C(x,y,z) = -\frac{1}{(x-y)(1-z)}\bigg\{ h_4(x) 
 - \frac{h_1(x) -  h_1(x,z)}{1-z}  - (x \leftrightarrow y) \bigg\} ,\notag \\
 &D^{RL3}_C(x,y) =  \frac{1}{2(1-y)}\bigg\{ h_5(x)
 -\frac{h_2(x) - h_2(x,y) }{1-y} \bigg\}  ,\notag 
 \end{align}
where $f_7(x)$, $f_8(x)$ are given by
\begin{align}
 f_7(x ) &= \frac{13-7x}{24(x-1)^3} -\frac{3+2x-2x^2}{12(x-1)^4}\log x , \notag \\
 f_8(x ) &= -\frac{1+5x}{8(x-1)^3} +\frac{x(2+x)}{4(x-1)^4}\log x . \notag
\end{align}
The loop functions for $C_{7\gamma }$ and $C_{8G}$ in (\ref{eq:chargino-right}): 
\begin{align}
  &D^{RR}_E(x,y) = \frac{1}{3(1-y)}\bigg\{ h_4(x) 
  - \frac{h_1(x)  - h_1(x,y) }{1-y} \bigg\}  \notag \\
  &~~~~~~~~~~~~~~~~~+ \frac{1}{2(1-y)}
  \bigg\{ f^{(1)}_7(x) -\frac{f^{(1)}_2(x) - f^{(1)}_2(x,y)}{1-y} \bigg\} ,\notag \\
 &D^{LR}_E(x) = -\frac{2}{3}f^{(0)}_1(x) + f^{(1)}_4(x) -\frac{1}{2}f^{(1)}_7(x)  ,\notag \\
 &D^{RR}_C(x,y) =\frac{1}{2(1-y)}\bigg\{ h_4(x)
  - \frac{h_1(x) - h_1(x,y)}{1-y} \bigg\} ,\notag \\
 &D^{LR}_C(x) = -f^{(0)}_1(x) + \frac{3}{2}f^{(1)}_4(x).\notag 
\end{align}


\end{document}